# Function and Form of Gestures in a Collaborative Design Meeting

Willemien Visser

LTCI, UMR 5141, CNRS - INRIA
46 rue Barrault, 75013 Paris, France
willemien.visser@TELECOM-ParisTech.fr

**Abstract.** This paper examines the relationship between gestures' function and form in design collaboration. It adopts a cognitive design research viewpoint. The analysis is restricted to gesticulations and emblems. The data analysed come from an empirical study conducted on an architectural design meeting. Based on a previous analysis of the data, guided by our model of design as the construction of representations, we distinguish representational and organisational functions. The results of the present analysis are that, even if form-function association tendencies exist, gestures with a particular function may take various forms, and particular gestural movements as regards form can fulfil different functions. Reconsidering these results and other research on gesture, we formulate the assumption that, if formal characteristics do not allow differentiating functional gestures in collaboration, context-dependent, semantic characteristics may be more appropriate. We also envision the possibility that closer inspection of the data reveal tendencies of another nature.

**Key words**: Gestural interaction. Cognitive design research. Collaborative design. Collaboration. Architectural design. Gesticulations. Emblems

## 1 Introduction

Dependent on researchers' aims, gestures have been analysed from several perspectives: for example, semiotic analysis [1], language development [2], human-computer interaction [3], gesture recognition and generation in interactive dialogue systems [4], and collaborative task-completion tools [5].

Working in the domain of cognitive design research [6], we are interested by the role of gesture in design collaboration. The aim of our research is double. Its main objective is an epistemic socio-cognitive psychology one—to understand the use of the different interaction modalities in professional collaboration. A long-term cognitive-ergonomics purpose is to contribute to the specification of remote collaborative-design environments, especially to facilitate the use of various semiotic modalities (multi-modal interaction) by designers working on remote locations.

In the study presented here, we examine if, in a collaborative design setting, gestures with a particular function have a particular form and/or if gestures with a particular form have a particular function. This question is relevant for several reasons. Data on the function-form relationship may present arguments for the debate concerning the idiosyncratic nature of gestures (cf. our Discussion and Conclusion). Further, such data may contribute to the specification of collaborative-design systems—or other multimodal interactive systems—in providing elements for the "translation" of functional, communicative, or semantic specifications into gestural movements [cf., for example, 7, but cf. also our Discussion and Conclusion].

Before introducing our analysis of the relationship between gestures' function and form, we shortly present our viewpoint on design and review previous research on gestural interaction in design collaboration. In the main section, we introduce the empirical study that provided the data that we use for our present analysis [8]. In the final section, we discuss further perspectives of this work.

## 2   Gestures in Design Collaboration

In this section, we introduce the viewpoint from which we analyse design and we review previous research on gestural interaction in design collaboration.

### 2.1   A Cognitive Perspective on Design

Design is an important, all-pervading domain of human activity. Not only are new sport cars and mobile phones the object of design, but so too are artefacts as diverse as traffic signals [9], route plans [10, 11], software [12], and, of course, buildings of all kinds [13].

This paper presents a study on architectural design, focusing on the types of socio-cognitive processes and structures implemented by designers collaborating on architectural projects.

In accordance with our theoretical framework, we consider designing as the construction of representations [6]. "Globally characterised, designing consists in specifying an artefact, *given requirements* that indicate—generally neither explicitly, nor completely—one or more functions to be fulfilled, and needs and objectives to be satisfied by the artefact, under certain conditions expressed by constraints. At a cognitive level, this specification activity consists in developing (generating, transforming, and evaluating) representations of the artefact *until they are so concrete, detailed, and precise* that the resulting representation—the *specifications* of the artefact—specify explicitly and completely the implementation of the artefact." [14, p. 117] Designing does not consist in *implementing* the specifications: it is not the fabrication of the artefact product, that is, in case of architectural design, the construction of the building. The representations that come out of architectural design are drawings and models of this artefact product. These representations are also artefacts, that is, entities created by people, "man-made as opposed to natural" [15].

"Cognitive design research" is the qualification for studies that examine design focusing on its cognitive aspects [6, 14]. Researchers from the engineering domain often use the term "design thinking" [16].

Design always involves several people—at least two, a client and a designer, or a designer and a user. Yet, given that design projects generally require the integration of information and knowledge from a variety of domains, they usually involve multiple competencies, and thus collaboration between people from different areas of expertise, working together in meetings. We consider as "designers" all those who, in such meetings, contribute to the specification of the artefact—even if their payslip may qualify them as, for example, "draftsman" or "programmer."

### 2.2 Gestural Interaction in Design Collaboration: Previous Studies

Compared with verbal and graphical interaction, gestural expression has barely been studied in collaborative design. In an analysis of the rare empirical studies on this use of gesture [17-19], we highlighted two functions [8]. (1) Gesture offers specific possibilities to render spatial (especially 3D) and motion-related qualities of entities, and to embody action sequences through their mimicked simulation. (2) Gesture plays an important organisational role.

In our previous research on gestures, we have developed a description language for graphico-gestural design activities [20]. Using this language to analyse the interaction in an architectural meeting, we have interpreted co-designers' graphico-gestural actions according to their functional roles in the project and in the meeting [21]. We also examined different forms of multimodal articulation between graphico-gestural and verbal modalities in parallel interactions between the designers, revealing alignment and disalignment between the designers regarding the focus of their activities [22, 23].

## 3 Analysing Function and Form of Gesture in a Design Meeting

This section presents the data we analysed to examine the function-form question, the functions of gestures we identified in a previous study, and our analysis of the relationship between function and form of gestures used in design collaboration.

### 3.1 Data: An Architectural Design Meeting

The data analysed come from the dataset for DTRS7 [The 7th Design Thinking Research Symposium, 24; DTRS7 dataset, P. Lloyd, J. McDonnell, F. Reid and R. Luck, 2007. These data are not publicly available for general distribution]. These data were provided to 24 researchers/research groups from different disciplines in order to confront, through different analyses of a same dataset, a representative variety of today's perspectives on design thinking. The dataset was made up of videos regarding naturally occurring design activity in the authentic setting of design practice. These were face-to-face, synchronous professional design meetings taking place in two different design firms (architecture and product design). We analysed the first architectural meeting, A1.

The A1 meeting took place in the pre-planning application stage of a project to design a new municipal crematorium with chapel, to be set in a landscaped site where existed already another crematorium. Data supplied were a video; a transcript of the audible part; plans at different scales, elevations, sketches, and orthographic projections referred to during the meeting; a 30-min video of an informal interview with the principal architect describing the background to the project. The video provided three views: top-view, medium-view, long-view (see Figure 1). The meeting took 2h 17min. The transcript had 2,342 transcript lines, corresponding to 987 speaker turns.

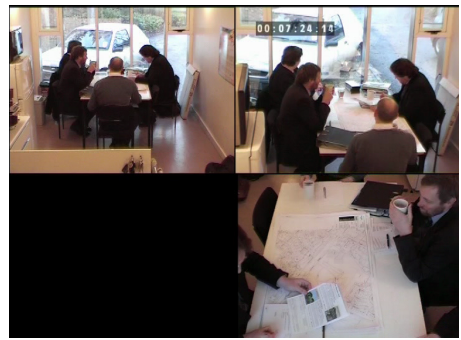

**Figure 1.** Three-view video of the A1 architectural meeting (from the DTRS7 dataset, 2007, P. Lloyd, J. McDonnell, F. Reid, and R. Luck)

The meeting involved three participants: Adam, the architect in charge of the project, and two clients, Anna, registrar of the cemetery, and Charles, an officer from local government representing the municipality's interests. A DTRS7 organiser (sitting at the end of the table) observed A1. Even if different roles can be distinguished among the three participants, we qualify them all as "designers," given our view of design as a cognitive activity, not necessarily the activity of somebody whose profession is "designer."

In our DTRS7 study [8], we analysed the entire meeting. However, we did not describe (as in the examples in Tables 1 and 2) all the gestures: our aim in this first study was to identify as much functions as possible. We therefore viewed the A1 video numerous times: during the first series of viewings, we noted and described an episode when we came across a gesture with a function not yet identified in our current analysis; afterwards, we went again through the video many times, searching for other instances of these functions.

We restricted our analysis to (1) gesticulations, that is, spontaneous, speech-accompanying gestures, and (2) emblems, quasi-linguistic, lexicalised gestures, with conventional forms and meanings, and which are not necessarily speech-accompanying [25, 26]. The qualification "gesture" is used for both in this paper.

In the DTRS7 study, we identified some 130 functional episodes (where an episode contains one or more gestures, see the two examples provided below in Tables 1 and 2, each one qualified "one episode"). We distinguished five main families of gestures, two of which each with two sub-families, two of which each again with their own two or three sub-families: representation (designation: identification, qualification, and comparison; specification), organisation (discourse and interaction management: management of one's own discourse, management of co-participants' interaction; functional design-action management), focalisation, modulation, and disambiguation. We did not retain all these distinctions for the analysis presented in this paper (see below).

Both in the DTRS7 and in the present analysis, we were the only judge for the identification of gestures, their function, and their form. In the identification of functions, we were guided by our model of design [6] (see below).

### 3.2 Functions of Gestures Used in Design Collaboration

We suppose that the activities that are performed or supported by gestures in design meetings fulfil one or more of the functions that cognitive design research has shown design activities to have [6]: (1) to contribute to the design *per se,* or (2) to have an organisational function.

The first family of activities—"representational gestures"—contributes to the construction of the representations that are to result in the specifications of the artefact. We indeed identified gestures that play a role in the generation, modification, or evaluation of such representations. Besides these representational specification gestures, we identified a second type of representational gestures, that is, representational designation gestures. Indeed, in addition to their classical function, that is, to point out an entity to one's co-participants, we observed that designational gestures can also have a distinctive design function: a designer can *design* an entity *through its designation* [8]. As regards data analysis, we wish to stress that it is difficult for an external observer (and for a designer's co-participants in a meeting) to distinguish between entities that "existed" already before their being designated and entities that are being designed through their designation. Designing is a continuous process: design working meetings generally are not events where a definitely finished project is being reported (and once the meeting over, design generally continues) [27] [cf. also 28's distinction between gestures that depict—iconic gestures—and gestures that conceive an entity].

Gestures also served organisational functions: the gestures of this second family—"organisational gestures"—contributed to the management of interaction [Bavelas' "interactive" gestures, 29] and to the organisation of functional design actions.

Relative to the DTRS7 study, we brought back to two main families (each with two subfamilies) the five families distinguished in our first analysis [8]. Identification, qualification, and comparison are no longer differentiated as separate functions. Management of one's own discourse and management of co-participants' interaction are both considered management of interaction now. Focalising, modulating, and disambiguating gestures are considered all three organisational now.

One may notice that the representational gestures in this study refer to a particular kind of representations. "Representational gestures" as identified by other authors [30, 31] receive this name because they "[represent] attributes, actions, or relationships of objects or characters" [30, p. 377], they "[represent] an aspect of the content of an utterance" [31, p. 160]. Here, the criterion for qualifying a gesture as "representational" is its contribution to the construction of a cognitive artefact, that is, a representation that is to result in the entity to be designed.

We chose the label "designational gestures" because it allows covering in an elegant way both the classical "deictics" or "pointing gestures" and the gestures that *design* through their designation.

### 3.3 The Relationship between Function and Form of Gesture in Design Collaboration

Do gestures with a particular function have a particular form, and do gestures with a particular form fulfil a particular function?

**Data Analysed**. To examine this question, we reanalysed the episodes of A1 that the DTRS7 study allowed identifying the functions of gesture in design collaboration, associating this time the form of the gestures to their function.

**Specifying Form.** Form can be specified and analysed at various levels (as function can be). We have to establish the relevant level(s). Most formal characterisations use rather low-level physical kinetic features, such as hand shape, palm orientation, movement, and location in gesture space. Our cognitive-ergonomics aim to contribute to the specification of collaborative-design environments is not yet formulated in terms of well-circumscribed specifications. We therefore adopt a medium or even high-level characterisation, using McNeill's [30] form-related subcategories for gesticulation: pointing, iconics/metaphorics [32], and beats. Rather than covering gestures with one particular form, these categories refer to families of forms: pointing gestures all share a movement towards the *designatum* [31]; iconics/metaphorics (re)present an object through the physical display of one or more of its characteristics; beats take the form of the hand beating time. If such broad categories allow for an association between gestures' function and form, a next step could be to examine the function-form relationship in terms of more low-level formal features (but cf. our Discussion and Conclusion section).

**Results.** We did not find a one-to-one function-form relationship.

*Gestures with a Particular Function do not Systematically have a Particular Form.* Representation was of course often performed using iconic/metaphoric gestures and emblems. However, it could also be through a pointing gesture. The episode presented in Table 1 gives an example of a pointing gesture, rather than an iconic/metaphoric gesture, that represents an entity. Through their designation, three possible itineraries are brought into life—that is, designed. Consecutively pointing with a finger three different routes on the plan, Adam indeed designs three possible itineraries for entering the cremator area. We will come back upon this analysis in the Discussion and Conclusion section.

**Table 1**. Designing three itineraries using pointing gestures
   Caption: The first description line indicates, between square brackets, the identity of the gesturer and the duration of the gesture by reference to that of the corresponding portion of the verbal protocol
   The second line presents a brief description of the gestural movement

| 34 | Adam | obviously there are numerous ways of getting into this accommodation |
|----|------|----------------------------------------------------------------------|
| 35 |      | that's route number one the second route is round the end of the pond [g_Ad...........................................……………………………... *tracing with a finger consecutively three lines over the plan* |
| 36 |      | and the third route is through the chapel so there's numerous ways of ...........................................……………….] |

There were nevertheless tendencies. Specification (the main representational design function) mostly occurred using iconic/metaphoric gestures. Designation in its classical signalling function (indicating something that already exists) was chiefly performed through pointing. By definition, designation in its design function was performed through pointing.

*Gestures with a Particular Form do not Systematically Fulfil a Particular Function.* We mentioned some tendencies above. Nevertheless, pointing had other uses than its signalling designation function; iconic/metaphoric gestures did not always specify design entities (cf. the example in Table 2). Pointing was also used to design (as illustrated in Table 1), to manage the interaction or design actions [cf. examples in 8], or to modulate components in one's discourse, another organisational function (cf. the example in Table 2).

Table 2 presents an episode in which Anna uses a variety of gestural movements to modulate elements of her discourse. It shows how Anna presents a book in which the author ("she" in line 34) mentions the already existing crematorium. When Anna wanted to order the book, she encountered many administrative problems, so she bought it herself. Telling this, she both highlights content elements and expresses their emotional loading, using different types of gestures.

- Pointing to the book, Anna uses this deictic gesture to highlight that "[they]'re mentioned in the book."
- Through long continuous hands opening, advancing gestures, she metaphorically emphasises discourse components.
- She also uses a sequence of detached short beats with this emphasis function (the "classic" use of beats).
- Enacting how a person holds a car steering wheel and then suddenly changes direction, Anna uses these iconic/metaphoric gestures to highlight her idea that people, normally, do not wish to visit cemeteries during their holidays.

**Table 2.** Modulating discourse components using various forms of gesture
Caption: *lchoag* = long continuous hands opening, advancing gestures
This table corresponds to Extract 6 from [8]

| 34 | Anna | yes she's been to have a look at our existing we're mentioned in the <br>                        [g_Anna.……………..<br>                       *points to the book* |
|---|---|---|
| 35 | | book quite a bit with the existing chapel she was quite impressed …......] |
| 36 | | because we've also got quite a lot of photographs and other things and <br> [g_Anna.…………………………]<br>*long continuous hands opening, advancing gestures* |
| 37 | | plans like the forward plan of extending the chapel originally the original <br>               [g_Anna.…………………………..…….]<br>              *long continuous hands opening, advancing gestures* |
| 38 | | idea so that was quite forward thinking in nineteen eighty or seventy |
| 39 | | whenever seventy eight when they decided on that + so I've sent off for |
| 40 | | that but you try and get them to raise a SAP order for it and you get how <br>     [g_Anna.…………………………..…..]         [g_Anna….…<br>    *lchoag*                                 *sequence of* |
| 41 | | often will we need the SPIRE BOOKS COMPANY I said well we'll never use <br> ………………………………………………]    [g_Anna………….]<br>*detached short beats*                               *lchoag* |
| 42 | | them again probably why do you need to do that oh for god's sake <br>                                  [g_Anna.……….]<br>                                  *lchoag* |
| 43 | | so I just went out and bought it and the one above as well a history <br>                        [g_Anna.……………..…..]<br>                         *lchoag* |
| 44 | | of cremations that's what we have at home on the bookshelf |
| 45 | | cheerful reading like you when you go on holiday with your part- your <br>                                        [g_An<br>                                       *lchoag* |
| 46 | | husband and wife + you see a sign for crematorium straight away <br> na……………..]                             [g_Anna.…...]<br>                                 *enacting how one* |
| 47 | | ignoring them whether you're abroad or in this country the <br>                     [g_Anna.…………………………….]<br>*holds a car steering wheel*<br>              *enacting how one suddenly changes direction* |
| 48 | | first thing you do straight away oh not another cemetery they go oh dear |

*The Multifaceted Nature of Gesture.* The results presented in the two previous sub-sections showed the multifaceted nature of gesture. Another example of this quality of gesture was that—as we already saw in these sub-sections—neither functional nor form categories were exclusive: a particular gesture could have various functions or combine various form-related characteristics. In A1, certain gestures used to specify could also serve focalisation, modulation, or disambiguation, that is, have an organisational function. The same observation was made for certain organisational gestures, which fulfilled other functions as well in a number of situations. As also mentioned already, gestures used to designate could have in addition an organisational function.

## 4  Discussion and Conclusion

We wish to highlight several outcomes with respect to the question examined in this paper and envision extensions of our study.

Our analysis did not allow attributing particular forms to particular functions—and v.v. A gesture with a particular function could take various forms, and a particular gestural movement as regards form could fulfil different functions. There were tendencies—designation being mostly performed through pointing gestures, and specification mainly through iconic/metaphoric gestures—but, as we mentioned, the distinctions between McNeill's [30] form-related categories are very global: a characterisation of the gestures in more low-level formal characteristics might not have shown such tendencies (but we are not going to examine this possibility, cf. below).

Another example of the multifaceted nature of gestures observed in the study was that functional and form categories were not exclusive. On the one hand, a particular gesture could fulfil different functions; on the other hand, one gesture could combine different form-related characteristics. Kendon [31] has also observed this in a detailed analysis of pointing (which he analyses as a function): pointing can be combined with other functions, for example, with a representational one (pp. 202-205). Nevertheless, Kendon notes as well that there are gestures that are "specialised" for pointing. Other authors have analysed as iconics, gestures that superficially may seem pointers (cf. our example in Table 1), observed that iconic gestures may be used as pointers [33], or shown that iconicity is not restricted to "iconic" gestures [34]. Doing so, they privilege other than formal features: they favour semantic or semiotic relations between the referent and its gestural expression.

If the formal characteristics we adopted do not allow differentiating functional gestures in collaboration, our next concern could be to find other features—formal or not—that allow such differentiation. Given our results—and, even more, on the basis of observations by other authors in the domain of gesture studies [35, 36]—we think that, rather than more low-level formal features, these will be context-dependent, semantic characteristics.

More tendencies than those identified here might exist—but especially tendencies of another than a formal nature seem to us more relevant at present. Many studies have shown indeed that formal features of gestural movements are not enough—and even not the most relevant attributes—to characterise gestures [7, 35, 36].

"Inter-speaker systematics" [35] has been searched for—and found—in several directions. "Representation techniques" [31] and "modes of representation" [34] are two related ones. Analysing iconicity (cf. above), Müller [34] identifies four different modes of representation that are used to "construct gestural meaning" (p. 321). In his detailed analysis of pointing gestures, Kendon [31] concludes that the specific form of these gestures "might vary systematically in relation to semantic distinctions of various sorts…. The form of the pointing gesture is not a matter of idiosyncratic choice or variation unrelated to the other things the speaker is doing." (p. 223) The author identifies seven different hand shape and hand shape/forearm orientation configurations.

Referring to Kendon [31], Bergmann and Kopp [35] analysed the use of iconic gestures in spatial descriptions. The authors examined the differential use of five representation techniques as a candidate for such "inter-speaker systematics." Analysing if there was a relationship between these representation techniques, Bergmann and Kopp observed inter-subjective correlations, but they also noticed individual differences (idiosyncratic aspects of gestures) [also underlined in 36].

To conclude, the study presented in this paper has illustrated again the complex relationship between gestures' function and form. It has pointed out some tendencies concerning form-function relationships, but we decided to not investigate further this specific issue. Instead, closer inspection of the data might reveal tendencies—or even regularities—of another nature (especially semantic and/or context-related).

**Acknowledgments**. The author wishes to thank an anonymous reviewer, whose questions, remarks, and suggestions in relation to an earlier version of this text have helped to advance her reflections on gesture and its relevant dimensions.